\documentclass[aps,nofootinbib,
preprint
]{revtex4}
\usepackage{amsmath}
\usepackage{amsfonts}
\usepackage{amssymb}
\usepackage{color}

\newcommand{\omits}[1]{}

\begin{document}


\title{Holographic Duals of Near-extremal Reissner-Nordstr{\o}m \\ Black Holes}

\author{Chiang-Mei Chen} \email{cmchen@phy.ncu.edu.tw}
\affiliation{Department of Physics and Center for Mathematics and Theoretical Physics,
National Central University, Chungli 320, Taiwan}

\author{Ying-Ming Huang} \email{y.m.huang26@gmail.com}
\affiliation{Department of Physics, National Central University, Chungli 320, Taiwan}

\author{Shou-Jyun Zou} \email{sgzou2000@gmail.com}
\affiliation{Department of Physics, National Central University, Chungli 320, Taiwan}

\date{\today}


\begin{abstract}
We consider the $\mathrm{AdS}_3/\mathrm{CFT}_2$ description of Reissner-Nordstr{\o}m black holes by studying their uplifted counterparts in five dimensions. Assuming a natural size of the extra dimension, the near horizon geometries for the extremal limit are exactly $\mathrm{AdS}_3 \times \mathrm{S}^2$. We compute the scattering amplitude of a scalar field, with a mode near threshold of frequency and extra dimensional momentum, by a near extremal uplifted black hole. The absorption cross section agrees with the two point function of the CFT dual to the scalar field.
\end{abstract}


\maketitle
\tableofcontents

\section{Introduction}
In the past years, there has been significant progress on elaborating the holographic principle \cite{'tHooft:1993gx, Susskind:1994vu,
Maldacena:1997re, Gubser:1998bc, Witten:1998qj} for the circumstances that do not rely on any fundamental theory, such as strings, by following an approach analogous to that studied in \cite{Brown:1986nw}. The main focus was on rotating black holes initiated from the Kerr/CFT correspondence \cite{Guica:2008mu}. The near horizon geometry of the extremal Kerr black hole consists of a warped $\mathrm{AdS}_3$ factor, and the left moving central charge of 2D CFT can be identified from the asymptotic symmetry group associated with appropriate boundary conditions. The Cardy formula for CFT entropy exactly reproduces the Bekenstein-Hawking entropy of the Kerr black holes. Soon after, similar calculations were applied to various different rotating black holes \cite{Hotta:2008xt, Lu:2008jk, Azeyanagi:2008kb, Chow:2008dp, Azeyanagi:2008dk, Nakayama:2008kg, Isono:2008kx, Peng:2009ty, Chen:2009xja, Loran:2009cr, Ghezelbash:2009gf, Lu:2009gj, Amsel:2009ev, Compere:2009dp, Krishnan:2009tj, Hotta:2009bm, Astefanesei:2009sh, Wen:2009qc, Azeyanagi:2009wf, Wu:2009di, Matsuo:2009sj, Peng:2009wx} and all found the remarkable agreement of CFT and black hole entropies. The essential further extension is to investigate the near extremal Kerr black holes in which right moving excitations of dual CFT are allowed. The relevant investigations used two different approaches. One way to probe the near extremal property is the process of near superradiant scattering of a scalar field (this could also be, more general, by the Dirac or gauge fields). The absorption cross section agrees with the two point function of the dual CFT operator \cite{Bredberg:2009pv, Hartman:2009nz, Cvetic:2009jn}. The other approach is the $\mathrm{AdS}_2/\mathrm{CFT}_1$ description \cite{Hartman:2008dq, Alishahiha:2008tv, Castro:2008ms} from the boundary stress tensor of the 2D effective action obtained by integrating out the angular coordinates \cite{Castro:2009jf}. Indeed, an expected promising scheme for having a better understanding of dual CFT (2D or 1D) is by constructing the corresponding stress tensor \cite{Dias:2009ex, Balasubramanian:2009bg, Amsel:2009pu, Castro:2009jf}.

It is then natural to question what is the dual CFT corresponding to non-rotating black holes, in particular, the spherical symmetric charged Reissner-Nordstr{\o}m (RN) black holes. It is well known that the near horizon geometry of a RN black hole is $\mathrm{AdS}_2 \times \mathrm{S}^2$ \cite{Bertotti:1959pf, Robinson:1959ev, Maldacena:1998uz}, therefore a holographic RN/CFT correspondence is expected. It is straightforward to compute the right moving central charge of the dual CFT by following the approach in \cite{Castro:2009jf} for the near extremal Kerr case. However, unlike the 2D CFT, the central charge of $\mathrm{CFT}_1$ depends on an ambiguous choice of normalization factor. A physically preferable choice for the normalization factor for RN/CFT was discussed in \cite{Chen:2009ht}, which gives a desirable value for the central charge. Moreover, the original method of deriving the left moving central charge for Kerr/CFT is essentially based on a fiberation expression of (warped) $\mathrm{AdS}_3$ as $\mathrm{AdS}_2 \times \mathrm{S}^1$. It turns out that the RN black hole can be consistently uplifted to one more higher dimension in a way that part of its U(1) gauge potential is transformed to the geometrical Kaluza-Klein (KK) vector field, which plays the role of the U(1) fiberation on the $\mathrm{AdS}_2$ basis. The uplifted counterpart was used to compute the left moving central charge in \cite{Hartman:2008pb, Garousi:2009zx}. Again, in this approach, there is a free parameter to be specified, namely the size of the extra dimension. In this paper, we will reexamine this issue and propose a natural choice for the radius of the extra dimensional circle in the way that the near horizon geometry is exactly $\mathrm{AdS}_3 \times \mathrm{S}^2$. This particular choice is compatible with the suggested normalization factor in the other approach \cite{Chen:2009ht}, and the left and right moving central charges, as expected, are identical, $c_L = c_R = 6 q^2 / G_4$.

The main goal of the paper is to verify the RN/CFT duality for the near extremal region following the method in \cite{Bredberg:2009pv}. We consider the scattering amplitude of a scalar field by a near extremal uplifted RN solution for a particular threshold limit that the frequency $\omega$ and momentum in the extra dimension $k$ satisfy specific conditions: $\omega - k$ is small but the corresponding right moving mode of CFT is finite. As we shall see later, this peculiar mode can probe the near horizon region to reveal certain information above the CFT. We found that the absorption cross section of the scattered scalar field really agrees with the two point function of the dual CFT operator. This agreement provides a strong evidence supporting the $\mathrm{AdS}_3/\mathrm{CFT}_2$ description for a RN black hole.

Evidently, the RN black holes can have both $\mathrm{AdS}_3/\mathrm{CFT}_2$ and $\mathrm{AdS}_2/\mathrm{CFT}_1$ descriptions. From the gravity point of view, the relation between the two AdS spaces is a dimensional reduction. In terms of field theory language, this relation indicates a light cone compatification between the two CFTs. Thus, one substantial importance of the RN/CFT correspondence is to provide a simple example from which we can directly examine the relation of $\mathrm{CFT}_2$ and $\mathrm{CFT}_1$ \cite{Gupta:2008ki}.

The outline of this paper is as follows. In Section II, we construct the uplifted RN black holes, derive the near horizon geometry, explore the dictionary for the variables of gravity and the CFT. The absorption cross section of a scalar wave in a near extremal uplifted RN black hole is computed in Section III and is compared with the CFT result in Section IV. We discuss the corresponding 4D perspectives in Section V, and finally conclude our results in Section VI. Moreover, the technique of dimensional reduction is briefly summarized in Appendix A.

\section{Uplifted RN Black Hole}
The 5D Einstein-Maxwell theory
\begin{equation}
I_5 = \frac1{16\pi G_5} \int d^5x \sqrt{- \hat g} \left( \hat R - \frac1{12} \hat F_{[3]}^2 \right),
\end{equation}
with the specific metric and form field assumptions (constant KK scalar field)
\begin{equation}
ds_5^2 = (dy + \mathcal{A}_\mu dx^\mu)^2 + ds_4^2, \qquad \hat A_{[2]} = \hat A \wedge dy,
\end{equation}
reduces to the 4D effective action
\begin{equation}
I_4 = \frac1{16\pi G_4} \int d^4x \sqrt{-g} \left( R - \frac14 \mathcal{F}^2 - \frac14 \hat F^2 \right).
\end{equation}
The extra dimension is supposed be a circle of radius $\ell$, i.e. $\hat y \sim \hat y + 2 \pi \ell$ and the two gravitational constants are related by $G_5 = 2 \pi \ell G_4$. The consistency of the constant KK scalar field assumption requires a relation between the two gauge potentials, i.e. $\hat A = \sqrt3 \mathcal{A}$, and consequently the 4D Einstein-Maxwell can be recovered (the technical details are summarized in Appendix A)
\begin{equation}
I_4 = \frac1{16\pi G_4} \int d^4x \sqrt{-g} \left( R - \frac14 F^2 \right),
\end{equation}
with
\begin{equation}
F^2 = \mathcal{F}^2 + \hat F^2 = 4 \mathcal{F}^2, \qquad \mathrm{or} \qquad \mathcal{A} = \frac12 A, \quad \hat A = \frac{\sqrt3}2 A.
\end{equation}
According to the reduction procedure, the 4D RN solution
\begin{eqnarray}\label{RN}
ds^2 &=& - f(r) dt^2 + \frac{dr^2}{f(r)} + r^2 d\Omega_2^2,
\nonumber\\
A &=& - \frac{2 q}{r} dt, \qquad f(r) = 1 - \frac{2m}{r} + \frac{q^2}{r^2},
\end{eqnarray}
where $m, q$ are mass and charge parameters, can be consistently uplifted to 5D spacetime with the KK vector potential $\mathcal{A} = - (q/r) dt$. The 5D counterpart solution, called the uplifted RN black hole, is
\begin{eqnarray}\label{RN5metric}
ds^2 &=& - f(r) dt^2 + \frac{dr^2}{f(r)} + r^2 d\Omega_2^2 + \left(dy - \frac{q}{r} dt \right)^2,
\nonumber\\
\hat A_{[2]} &=& - \sqrt3 \frac{q}{r} \, dt \wedge dy.
\end{eqnarray}
It is worth noting that this simple uplifting procedure cannot accommodate dyonic RN black holes because the appearance of the Chern-Simons correction will turn on a 3-form field strength in the 4D reduced action. A new feature of the uplifted RN black hole is the emergence of the ergosphere at $- g_{tt} = 1 - 2 m /r = 0$ which seems to hint the existence of superradiant mode.

The near horizon geometry of the near extremal uplifted RN black hole can be obtained by introducing a new coordinate, $\chi$, normalized with period of $2\pi$:
\begin{equation}
\chi = \frac1{\ell} (y - t), \quad \mathrm{or} \quad y = \ell \chi + t,
\end{equation}
and taking the limit
\begin{equation}
r \to q + \epsilon \rho, \qquad t \to \frac1{\epsilon} q^2 \tau, \qquad m \to q + \epsilon^2 \frac{\rho_0^2}{2q},
\end{equation}
which leads to the near horizon solution
\begin{eqnarray} \label{NHRN5}
ds^2 &=& q^2 \left[ - (\rho^2 - \rho_0^2) d\tau^2 + \frac{d\rho^2}{\rho^2 - \rho_0^2} + d\Omega_2^2 + \frac{\ell^2}{q^2} \left(d\chi + \frac{q}{\ell} \rho d\tau \right)^2 \right],
\nonumber\\
\hat A_{[2]} &=& \sqrt3 q \ell \rho \, d\tau \wedge d\chi.
\end{eqnarray}
After taking the near horizon limit, the inner/outer horizon radius, Hawking temperature, and black hole entropy become
\begin{eqnarray}
r_\pm &=& q \pm \epsilon \rho_0,
\nonumber\\
T_H &=& \frac1{4\pi} \frac{r_+ - r_-}{r_+^2} = \epsilon \frac{\rho_0}{2 \pi q^2},
\nonumber\\
S_{BH} &=& \frac{A_5}{4 G_5} = \frac{A_4}{4 G_4}= \frac{\pi}{G_4} (q^2 + 2 \epsilon q \rho_0).
\end{eqnarray}

The size of the extra dimension $\ell$ in principle is a free parameter. However, from the near horizon metric (\ref{NHRN5}), there is an obvious simple and natural choice of $\ell = q$ and the near horizon geometry is then exactly $\mathrm{AdS}_3 \times \mathrm{S}^2$. The $\mathrm{AdS}_3$ space has two $SL(2,R)$ symmetries associated with the $\tau$-translation (right sector) and $\chi$-translation (left sector). By the geometric relations of the coordinates ($t, y$) and ($\tau, \chi$), we can identify the left and right moving modes, $n_L$ and $n_R$, from the frequency $\omega$ and $y$-direction momentum $k$ as
\begin{equation}
- i \omega t + i k y = - i n_R \tau + i n_L \chi \quad \Rightarrow \quad n_L = k q, \quad n_R = \frac{q^2(\omega - k)}{\epsilon}.
\end{equation}
The right moving mode is finite only when $\omega - k$ is small, of order $\epsilon$ corresponding to a threshold bound which is capable of probing a certain nature of the dual CFT. The next step of building up the gravity-CFT dictionary is to compare the Boltzmann factors. On the gravity side, one needs a time-like Killing vector which is well-defined near the horizon. It is clear that $\partial_t$ is not the correct choice since it becomes space-like inside the ergosphere. A suitable time-like Killing vector is $\partial_t + \Phi_e \partial_y$ where $\Phi_e$ is just the chemical potential of the 4D RN back holes
\begin{equation}
\Phi_e = \frac{q}{r_+} = 1 - \epsilon \frac{\rho_0}{q}.
\end{equation}
Then the temperatures of the CFT can straightforwardly be read out from the Boltzmann factors:
\begin{equation}
- \frac{\omega - \Phi_e k}{T_H} = - \frac{n_L}{T_L} - \frac{n_R}{T_R} \qquad \Rightarrow \qquad T_R = \frac{\rho_0}{2\pi}, \quad T_L = \frac1{2\pi}.
\end{equation}


\section{Macroscopic Greybody Factor}
Consider a scalar field of mass $\mu$ propagating in the background of (\ref{RN5metric}), the Klein-Gordon (KG) equation
\begin{equation} \label{KG}
\nabla_\alpha \nabla^\alpha \Phi - \mu^2 \Phi = 0,
\end{equation}
can be simplified by assuming the following form of the scalar field
\begin{equation} \label{APhi5}
\Phi(t, r, \theta, \phi, y) = \mathrm{e}^{-i \omega t + i n \phi + i k y} S(\theta) R(r),
\end{equation}
and reduces to two decoupled equations by separation of variables:
\begin{eqnarray}
\partial_r (\Delta \partial_r R) + \left[ \frac{(\omega r - k q)^2 r^2}{\Delta} - (\mu^2 + k^2) r^2 - \lambda_l \right] R &=& 0,
\label{EqRr} \\
\frac1{\sin\theta} \partial_\theta (\sin\theta \partial_\theta S_l) + \left( \lambda_l - \frac{n^2}{\sin^2\theta} \right) S_l &=& 0,
\label{EqS}
\end{eqnarray}
where $\Delta = r^2 f = (r - r_+) (r - r_-)$, the separation constant takes the standard value $\lambda_l = l (l + 1)$, and the solutions for $S_l$ are just the standard spherical harmonic functions.

To simplify the radial equation, we introduce a new dimensionless coordinate
\begin{equation}
z = \frac{r - r_+}{r_+}, \qquad r = r_+ (z + 1),
\end{equation}
consequently leading to the following relations:
\begin{equation}
\Delta = r_+^2 z (z + z_0), \qquad z_0 = \frac{r_+ - r_-}{r_+} = \epsilon \frac{2\rho_0}{q}, \qquad r_+ = \frac{q}{\sqrt{1 - z_0}}.
\end{equation}
Finally, the radial equation can formally be written as
\begin{equation} \label{EqRz}
\partial_z \left[ z (z + z_0) \partial_z R \right] + V(z) R = 0,
\end{equation}
where the potential is given by
\begin{equation}
V = \frac{q^2 (z + 1)^2}{1 - z_0} \left[ \frac{\left[ \omega (z + 1) - k \sqrt{1 - z_0} \right]^2}{z (z + z_0)} - \mu^2 - k^2 \right] - \lambda_l.
\end{equation}
However, it is not possible to obtain the general solution of equation (\ref{EqRz}). In order to retrieve the absorption cross section, we should solve this equation in the near region $z \ll 1$ and at the far region $z \gg z_0$ respectively. For the near extremal limit, $z_0 \to 0$, there is an overlap region, $z_0 \ll z \ll 1$, which allows to relate the integration constants in the solutions of two different regions.

\subsection{Near region}
For the near region, $z \ll 1$, and for the near extremal limit, $z_0 \to 0$, the potential truncates to
\begin{eqnarray}
V_\mathrm{near} &=& q^2 \left[ \frac{\left[ (2 \omega - k) z + \omega - k + \frac12 \omega z_0 \right]^2}{z (z + z_0)} + 2 \omega (\omega - k) - \mu^2 - k^2 \right] - \lambda_l
\\
&\simeq& q^2 \left[ \frac{k^2 z_0^2 }{4 z (z + z_0)} - \mu^2 \right] - \lambda_l.
\nonumber
\end{eqnarray}
Hereafter, the notation $\simeq$ means after taking the threshold limit $\omega - k = \epsilon \, n_R /q^2$. The radial equation then reduces to
\begin{equation}
\partial_z \left[ z (z + z_0) \partial_z R \right] + V_\mathrm{near} R = 0,
\end{equation}
and two independent solutions corresponding to ingoing and outgoing waves are
\begin{eqnarray}
R_\mathrm{near}^\mathrm{(in)}(z) \!&\!=\!&\! C_0^\mathrm{(in)} \! z^{-i b} \left( 1 \!+\! \frac{z}{z_0} \right)^{i (b - a)} \! {}_2F_1\!\left( \frac12 \!+\! \beta \!-\! ia, \frac12 \!-\! \beta \!-\! ia; 1 \!-\! i 2 b; -\frac{z}{z_0} \right),
\\
R_\mathrm{near}^\mathrm{(out)}(z) \!&\!=\!& \! C_0^\mathrm{(out)} \! z^{i b} \left( 1 \!+\! \frac{z}{z_0} \right)^{i (b - a)} \! {}_2F_1\!\left( \frac12 \!-\! \beta \!+\! i(2b \!-\! a), \frac12 \!+\! \beta \!+\! i(2b \!-\! a); 1 \!+\! i 2 b; \!-\! \frac{z}{z_0} \right)\!,
\end{eqnarray}
where the constants $a, b$ and $\beta$ are are defined by
\begin{eqnarray}
a &=& q (2 \omega - k) \simeq q k,
\\
b &=& \frac12 \omega q + \frac{q (\omega - k)}{z_0} \simeq \frac12 q k + \frac{n_R}{2 \rho_0},
\\
\beta &=& \sqrt{\frac14 + \lambda_l + q^2 (\mu^2 - 6 \omega^2 + 6 \omega k)} \simeq \sqrt{\frac14 + \lambda_l + q^2 \mu^2}.
\label{defbeta}
\end{eqnarray}
These three parameters are principle elements for comparison with the dual CFT results. The value of $\beta$ is always real.

We are interested only in the ingoing wave. At the limit near the horizon, $z \to 0$, we have
\begin{equation} \label{nearin}
R_\mathrm{near}^\mathrm{(in)}(z) \approx C_0^\mathrm{(in)} \, z^{-i b},
\end{equation}
and at the limit close to the far region, namely $z \gg 1$, it behaves like
\begin{equation} \label{nearfar}
R_\mathrm{near}^\mathrm{(in)} \to A_0^\mathrm{(in)} z^{-\frac12 + \beta} + B_0^\mathrm{(in)} z^{-\frac12 - \beta},
\end{equation}
where
\begin{eqnarray} \label{ABC}
A_0^\mathrm{(in)} &=& C_0^\mathrm{(in)} (z_0)^{\frac12 - \beta - ib} \frac{\Gamma(1 - i 2 b) \Gamma(2 \beta)}{\Gamma(\frac12 + \beta - ia) \Gamma(\frac12 + \beta - i (2b - a))},
\nonumber\\
B_0^\mathrm{(in)} &=& C_0^\mathrm{(in)} (z_0)^{\frac12 + \beta - ib} \frac{\Gamma(1 - i 2 b) \Gamma(-2 \beta)}{\Gamma(\frac12 - \beta - ia) \Gamma(\frac12 - \beta - i (2b - a))}.
\end{eqnarray}

\subsection{Far region}
For the far region, $z \gg z_0$, the potential truncates to
\begin{eqnarray}
V_\mathrm{far} &=& q^2 (\omega z + 2 \omega - k)^2 + 2 q^2 \omega (\omega - k) - q^2 (\mu^2 + k^2) (z + 1)^2 - \lambda_l
\\
&\simeq& q^2 k^2 (z + 1)^2 - q^2 (\mu^2 + k^2) (z + 1)^2 - \lambda_l,
\end{eqnarray}
and the radial equation reduces to
\begin{equation}
\partial_z (z^2 \partial_z R) + V_\mathrm{far} R = 0.
\end{equation}
Two independent solutions can be obtained in terms of Kummer functions:
\begin{eqnarray}
R_\mathrm{far}^\mathrm{(in)}(z) &=& C_\infty^\mathrm{(in)} \, z^{- \frac12 + \beta} \mathrm{e}^{i \zeta z} \; M\left(\frac12 + \beta + i \kappa; 1 + 2 \beta; - i 2 \zeta z \right),
\\
R_\mathrm{far}^\mathrm{(out)}(z) &=& C_\infty^\mathrm{(out)} \, z^{- \frac12 + \beta} \mathrm{e}^{i \zeta z} \; U\left(\frac12 + \beta + i \kappa; 1 + 2 \beta; - i 2 \zeta z \right),
\end{eqnarray}
where $\beta$ was given in (\ref{defbeta}) and the other two parameters are defined as
\begin{equation} \label{zeta}
\zeta = q \sqrt{\omega^2 - k^2 - \mu^2}, \qquad \kappa = \frac{q^2 (2 \omega^2 - \omega k - k^2 - \mu^2)}{\zeta}.
\end{equation}
The ingoing and outgoing waves can be determined by using the asymptotic behavior of the Kummer functions \cite{AbSt64}. In order to approach the threshold limit $\omega - k \to \epsilon$ but still to ensure the propagating waves, i.e. $\zeta$ be real, the scalar field should be massless, $\mu = 0$. Nevertheless, the the scalar waves always have enormously long wavelength near the threshold limit. Moreover, over the threshold limit, $\omega < k$, the wave number $\zeta$ becomes imaginary and wave solutions fade out.

In order to match with the near region result, we consider the limit $z \to 0$
\begin{equation} \label{farnear}
R_\mathrm{far} \approx C_\infty^\mathrm{(in)} \, z^{- \frac12 + \beta} + C_\infty^\mathrm{(out)} \frac{\Gamma(2 \beta)}{\Gamma(\frac12 + \beta + i \kappa)} \, (- i 2 \zeta)^{-2\beta} \, z^{- \frac12 - \beta}.
\end{equation}
For computing the absorption cross section, we need the asymptotic behavior of the propagating waves
\begin{equation} \label{farasy}
R_\mathrm{far} \to C_\infty^\mathrm{(in)} \frac{\Gamma(1 + 2 \beta)}{\Gamma(\frac12 + \beta + i \kappa)} (- i 2 \zeta)^{- \frac12 - \beta + i \kappa} \, \mathrm{e}^{- i \zeta z} \, z^{- 1 + i \kappa} + C_\infty^\mathrm{(out)} (- i 2 \zeta)^{-\frac12 - \beta - i \kappa} \, \mathrm{e}^{i \zeta z} \, z^{- 1 - i \kappa}.
\end{equation}

\subsection{Near-far matching}
The results (\ref{nearfar}) and (\ref{farnear}) should match at the overlap region, $z_0 \ll z \ll 1$, which gives the relations of the coefficients:
\begin{equation} \label{match}
C_\infty^\mathrm{(in)} = A_0^\mathrm{(in)}, \qquad C_\infty^\mathrm{(out)} = B_0^\mathrm{(in)} (- i 2 \zeta)^{2\beta} \frac{\Gamma(\frac12 + \beta + i \kappa)}{\Gamma(2 \beta)}.
\end{equation}
Moreover, the powers of $z$ in (\ref{nearfar}, \ref{farnear}) add more ``words'' to the gravity-CFT dictionary, namely that the conformal dimensions of the associated CFT operators should be
\begin{equation}
h_L = h_R = \frac12 + \beta.
\end{equation}

\subsection{Absorption cross section}
According to the flux expression of the scalar field
\begin{equation}
\mathfrak{F} = \frac{2\pi}{i} \left( \Phi^* \Delta \partial_r \Phi - \Phi \Delta \partial_r \Phi^* \right),
\end{equation}
we can compute the flux absorbed by the black hole by considering (\ref{nearin}),
\begin{equation}
\mathfrak{F}_\mathrm{abs} = - \left| C^{(\mathrm{in})}_0 \right|^2 \; 4 \pi b z_0 r_+,
\end{equation}
and the incident flux from infinity by taking ingoing part of (\ref{farasy}),
\begin{equation}
\mathfrak{F}_\mathrm{in} = -\left| C^{(\mathrm{in})}_\infty \right|^2 \frac{|\Gamma(1 + 2\beta)|^2}{|\Gamma(\frac12 + \beta + i \kappa)|^2} \, \mathrm{e}^{\kappa \pi} (2 \zeta)^{-\frac12 - 2\beta} \, \left( 4 \pi \zeta r_+ \right).
\end{equation}
Here the minus signs indicate the feature of ingoing flux. The absorption cross rate of the radial flux is
\begin{eqnarray} \label{sigma}
\sigma_\mathrm{abs} &=& \frac{\mathfrak{F}_\mathrm{abs}}{\mathfrak{F}_\mathrm{in}} = \frac{2 b z_0 \, (2 \zeta)^{-\frac12 + 2\beta} \left| \Gamma(\frac12 + \beta + i \kappa) \right|^2}{\mathrm{e}^{\kappa \pi} \left| \Gamma(1 + 2 \beta) \right|^2} \; \frac{\left| C^{(\mathrm{in})}_0 \right|^2}{\left| C^{(\mathrm{in})}_\infty \right|^2}
\\
&=& \frac{(2 \zeta)^{-\frac12 + 2\beta} z_0^{2\beta} \left| \Gamma(\frac12 \!+\! \beta \!+\! i \kappa) \right|^2 \sinh(2 \pi b)}{\pi \mathrm{e}^{\kappa \pi} \left| \Gamma(1 + 2\beta) \right|^2 \left|\Gamma(2 \beta) \right|^2}  \left| \Gamma\left( \frac12 \!+\! \beta \!-\! i a \right) \right|^2 \left|\Gamma\left( \frac12 \!+\! \beta \!-\! i (2b \!-\! a) \right) \right|^2. \nonumber
\end{eqnarray}
Here the ratio of $| C^{(\mathrm{in})}_0 |^2 / | C^{(\mathrm{in})}_\infty |^2$ is determined by the matching (\ref{match}) and also by the relation (\ref{ABC}). Moreover, the property of the Gamma function $|\Gamma(1 - i2b)|^2 = 2 b \pi/\sinh(2b\pi)$ has been used. An important remark is that the superradiance occurs when the value of parameter $b$ is negative, i.e. $\omega < k /(1 + \frac12 z_0)$. The 4D description of this result will be given in the Section~\ref{4DP}.

\section{Microscopic Greybody Factor}
The two-point functions in 2D CFT are determined by conformal invariance. The general expression for the absorption cross section with respect to an operator of (left, right) dimension ($h_L, h_R$), charge ($q_L, q_R$) and momenta ($\omega_L, \omega_R$) at temperature ($T_L, T_R$) and chemical potential ($\mu_L, \mu_R$) was derived in \cite{Bredberg:2009pv, Hartman:2009nz}:
\begin{equation}\label{CFTp}
P_\mathrm{abs} \sim T_L^{2h_L - 1} T_R^{2h_R - 1} \left( \mathrm{e}^{\pi (\tilde\omega_L + \tilde\omega_R)} \pm \mathrm{e}^{- \pi (\tilde\omega_L + \tilde\omega_R)} \right) \left| \Gamma(h_L + i \tilde\omega_L) \right|^2 \, \left| \Gamma(h_R + i \tilde\omega_R) \right|^2,
\end{equation}
where
\begin{equation}
\tilde\omega_L = \frac{\omega_L - q_L \mu_L}{2\pi T_L}, \qquad \tilde\omega_R = \frac{\omega_R - q_R \mu_R}{2\pi T_R}.
\end{equation}
The CFT absorption cross section matches with the dual gravitational result (\ref{sigma}) after identifying
\begin{equation}\label{Iden}
\tilde\omega_L = a, \qquad \tilde\omega_R = 2 b - a,
\end{equation}
and then the combination of the two exponential terms, with minus sign, produces the hyperbolic sine function. Moreover, the contribution of $T_L$ in (\ref{CFTp}) precisely reproduces the $z_0$ part in (\ref{sigma}). Unlike the Kerr/CFT, the dual operator corresponding to the scalar field in a near extremal RN black hole is much simpler with $q_L = q_R = 0$, $(\omega_L, \omega_R) = (n_L, n_R)$ and $\mu_L = \mu_R = 0$. One can easily check that the identifications (\ref{Iden}) hold for the near threshold mode.

\section{4D Perspectives} \label{4DP}
The KG equation for a charged scalar field with mass $\bar\mu$ and charge $e$
\begin{equation}
\Phi(t, r, \theta, \phi) = \mathrm{e}^{-i \omega t + i n \phi} S(\theta) R(r),
\end{equation}
in the background of a 4D RN black hole decouples to two equations:
\begin{eqnarray}
\partial_r (\Delta \partial_r R) + \left[ \frac{(\omega r - 2 e q)^2 r^2}{\Delta} - \bar\mu^2 r^2 - \lambda_l \right] R &=& 0,
\\
\frac1{\sin\theta} \partial_\theta (\sin\theta \partial_\theta S_l) + \left( \lambda_l - \frac{n^2}{\sin^2\theta} \right) S_l &=& 0.
\end{eqnarray}
The angular equation is exactly same as (\ref{EqS}). Moreover, the radial equation is also identical with (\ref{EqRr}) when the parameters are related by
\begin{equation}
e = \frac12 k, \qquad \bar\mu^2 = \mu^2 + k^2.
\end{equation}
Therefore, the 5D momentum mode $k$ corresponds to the charge of the 4D scalar. The factor half is due to the $k$-momentum coupling only with the KK vector portion, namely half of the original 4D RN background charge. Moreover, the excitation of the extra momentum also enlarges the mass of the 4D scalar field. Regarding the near threshold mode with $\mu = 0$, from the 4D point of view, it implies a special limit of $\bar\mu = 2 e$ and $\omega - \bar\mu \to 0$.

The superradiance condition for a charged scalar field in RN black hole is $\omega < e q / r_+$ \cite{Nakamura:1976nc}. For the near extremal limit, $z_0 \to 0$, it reduces to $\omega < k \sqrt{1 - z_0}$, which agrees with the condition of negative $b$. Moreover, for 4D effective scalar field, the mass is always greater than $k$ (i.e. charge). This is the reason why the wave number in the solution of radial part of the scalar field, equivalently the $\zeta$ in (\ref{zeta}), turns into imaginary over the threshold limit when $\omega < k$.

\section{Conclusion}
In this paper, we investigated the $\mathrm{AdS}_3/\mathrm{CFT}_2$ description for the near extremal RN black holes by considering their uplifted counterparts. Unlike the Kerr spacetimes, although an ergosphere is formed in uplifted RN black holes, yet the backgrounds, in the near extremal limit, do not allow superradiant modes for scalar fields because the wave number turns out to be imaginary. Nevertheless, for an exceptional threshold limit the scalar waves can be used to probe of the near horizon region and can reveal convincing properties which extend the confirmation of RN/CFT correspondence to the near extremal limit.

One promising feature of the RN/CFT is that the uplifted RN black holes can provide simpler backgrounds for investigating the holographic correspondence. The near horizon geometry contains an exact $\mathrm{AdS}_3$ space instead of warped $\mathrm{AdS}_3$, and accordingly the dual CFT operators are much simpler compared with Kerr/CFT. The factor $\beta$ in the conformal weight of the dual operator is always real. It is very interesting to study the connection between the $\mathrm{AdS}_3/\mathrm{CFT}_2$ and $\mathrm{AdS}_2/\mathrm{CFT}_1$ descriptions by considering the RN black holes. From the gravity side, the relation of $\mathrm{AdS}_3$ and $\mathrm{AdS}_2$ is more clear as a dimensional reduction. We expect that this relation can reveal the connection of $\mathrm{CFT}_2$ and $\mathrm{CFT}_1$ via holographical duality.

\begin{appendix}

\section{Dimensional Reduction}
In this Appendix, we briefly summarize the KK reduction technique \cite{Stelle:1998xg}. The special class of solutions in the ($D+1$) dimensional gravitational theory coupled to a scalar field and a $n$-form field,
\begin{equation}
\hat I = \int d^{D+1}x \sqrt{-\hat g} \left[ \hat R - \frac12 (\hat\nabla \phi)^2 - \frac1{2 \cdot n!} \mathrm{e}^{\hat a \phi} \hat F_{[n]}^2 \right],
\end{equation}
with a space-like Killing vector, $\partial_y$, is effectively equivalent to a $D$ dimensional effective action derived by imposing the following ansatz for the metric and form-field
\begin{equation}
ds_{D+1}^2 = \mathrm{e}^{2 \alpha \varphi} ds_D^2 + \mathrm{e}^{2 \beta \varphi} (dy + \mathcal{A}_\mu dx^\mu)^2, \qquad \hat A_{[n-1]} = A_{[n-1]} + A_{[n-2]} \wedge dy.
\end{equation}
The coefficients $\alpha$ and $\beta$ are fixed by (i) $\beta = - (D - 2) \alpha$ to ensure that the reduced action is in the Einstein frame, and (ii) $\alpha^{-2} = 2 (D - 1)(D - 2)$ for a conventionally normalized kinetic term of $\varphi$. Then the KK reduced action is
\begin{eqnarray}
I &=& \int d^Dx \sqrt{- g} \left[ R - \frac12 (\nabla \phi)^2 - \frac12 (\nabla \varphi)^2 - \frac14 \mathrm{e}^{-2 (D-1) \alpha \varphi} \mathcal{F}_{[2]}^2 \right.
\nonumber\\
&& \qquad \left. - \frac1{2 \cdot n!} \mathrm{e}^{-2 (n-1) \alpha \varphi + \hat a \phi} \hat F_{[n]}^{\prime2} - \frac1{2 \cdot (n-1)!} \mathrm{e}^{2 (D-n) \alpha \varphi + \hat a \phi} \hat F_{[n-1]}^2 \right],
\end{eqnarray}
where $\hat F'_{[n]} = \hat F_{[n]} - \hat F_{[n-1]} \wedge \mathcal{A}$, including a Chern-Simons correction.

For the purpose of the uplifted 4D Einstein-Maxwell solution we focus on the case $\hat a = 0, \phi = 0, D = 4$ corresponding to $\alpha^{-1} = 2\sqrt3$.  There are two possible ways to get a 4D U(1) gauge field, namely either $n = 2$ or $n = 3$. The first possibility is $n = 2$ and the $\varphi = 0$ truncation requires $\hat F_{[1]} = 0$ and $\hat F_{[2]}^{\prime2} = \hat F_{[2]}^2 = - 3 \mathcal{F}_{[2]}^2$. However, the 2-form field $\mathcal{F}_{[2]}$ has an opposite sign of its square, so only its Hodge dual can give the desirable sign. The other possible solution is $n = 3$ and the vanishing KK scalar condition requires $\hat F_{[3]} = 0$ and $\hat F_{[2]}^2 = 3 \mathcal{F}_{[2]}^2$. Thus the 4D action is
\begin{equation}
I = \int d^4x \sqrt{-g} \left( R - \frac14 \mathcal{F}_{[2]}^2 - \frac14 \hat F_{[2]}^2 \right) = \int d^4x \sqrt{-g} \left( R - \mathcal{F}_{[2]}^2 \right),
\end{equation}
consistently truncated to 4D Einstein-Maxwell theory by assuming $\mathcal{A} = \frac12 A, \hat A = \frac{\sqrt3}2 A$.

\end{appendix}

\section*{Acknowledgement}
We would like to thank A. Ishibashi, H.~Kodama, H.~Lu, J.~M.~Nester, J.~R.~Sun and especially W.~Song for very valuable discussions. CMC is grateful to the KEK for its hospitality in the revising stage of this paper. This work was supported by the National Science Council of the R.O.C. under the grant NSC 96-2112-M-008-006-MY3 and in part by the National Center of Theoretical Sciences (NCTS).


\end{document}